\pdfoutput=1
\documentclass[prl,showpacs,twocolumn]{revtex4-1}
\usepackage{graphicx}
\usepackage{amsmath}
\usepackage{amssymb}
\usepackage{xspace}

\begin{document}

\title{Glassy behavior in a binary atomic mixture}

\author{Bryce Gadway}
\email{bgadway@ic.sunysb.edu}
\author{Daniel Pertot}
\thanks{Present address: AMOP Group, Cavendish Laboratory, University of Cambridge, Cambridge, United Kingdom}
\author{Jeremy Reeves}
\author{Matthias Vogt}
\author{Dominik Schneble}
\affiliation{Department of Physics and Astronomy, Stony Brook University,
Stony Brook, NY 11794-3800, USA}
\date{\today}

\begin{abstract}
We experimentally study one-dimensional, lattice-modulated Bose gases in the
presence of an uncorrelated disorder potential formed by localized impurity
atoms, and compare to the case of correlated quasi-disorder formed by an
incommensurate lattice. While the effects of the two disorder realizations
are comparable deeply in the strongly interacting regime, both showing
signatures of Bose glass formation, we find a dramatic difference near the
superfluid-to-insulator transition. In this transition region, we observe
that random, uncorrelated disorder leads to a shift of the critical lattice
depth for the breakdown of transport as opposed to the case of correlated
quasi-disorder, where no such shift is seen. Our findings, which are
consistent with recent predictions for interacting bosons in one dimension,
illustrate the important role of correlations in disordered atomic systems.
\end{abstract}

\maketitle

The presence of disorder is inherent to solid state systems, and it has
profound effects on transport in a variety of contexts, ranging from electron
conductivity in metals to dirty superconductors~\cite{Anderson-DirtySC}.
Quantum gases in optical lattices~\cite{BlochDalibardZwergerReview-2008} can,
with a high degree of experimental control, elucidate the role played by
disorder in a number of physical phenomena. Recently, Anderson localization
of matter-waves in disordered potentials has been observed for
non-interacting
gases~\cite{Billy-AndersonLocalization-2008,Roati-AndersonLocalization-2008},
and additionally, the reemergence of superfluidity due to repulsive
interactions~\cite{Deissler-DisorderWithInteractions-2010}. Discerning the
(at times) competing roles of disorder and interactions is key to the
understanding of Bose-glass
behavior~\cite{Giamarchi-BoseGlass-1988,Fisher-Bosons-SF-IN-1989,Krauth-SF-IN-DisorderBosons-1991} in
strongly interacting disordered systems. Ultracold atomic
systems~\cite{Fallani-TowardsBG-2007,White-SpeckleDisorder-2009,*Pasienski-Disorder-2010}
should provide a versatile testbed to aid in this
endeavor~\cite{Sanchez-Palencia-Lewenstein-NP-2010}.

Previous studies of ultracold atoms in disordered potential landscapes,
generated by optical fields, have generally suffered from strong correlations
of the disorder that decay over length scales greater than either the healing
length of the superfluid or the lattice spacing. This is true for both
speckle
potentials~\cite{Billy-AndersonLocalization-2008,White-SpeckleDisorder-2009,*Pasienski-Disorder-2010}
that are diffraction-limited to structures on the order of the generating
laser field's wavelength, and quasi-disordered bichromatic
lattices~\cite{Fallani-TowardsBG-2007,Roati-AndersonLocalization-2008,Deissler-DisorderWithInteractions-2010},
which over large distances exhibit perfect correlations that may make them
rather unsuitable for the realization of true disorder. To circumvent these
limitations, recently it has been
proposed~\cite{Gavish-2005-DisorderAtoms,Paredes-QuantumParallelism-2005,RoscildeQuantumEmulsion-2007,Horstmann-PRL-2010}
to use atomic impurities, which can be confined to regions much smaller than
a lattice spacing, to act as point-like defects.

The sudden quench of an atomic impurity field acting on mobile particles has
been proposed~\cite{Paredes-QuantumParallelism-2005,Horstmann-PRL-2010} for
the study of dynamical, out-of-equilibrium response to disorder.
Alternatively, theoretical
studies~\cite{RoscildeQuantumEmulsion-2007,Buonsante-BosonMixtures-2008} have
shown that even a slow ``freeze-out'' of the tunneling of one species from an
initially homogenous mixture can lead to metastable ``quantum emulsion''
states, typified by local separation between frozen and mobile atoms (for
repulsive interactions) and displaying properties similar to an equilibrium
Bose glass. The study of quantum emulsions may help shed light on the
coherence-loss mechanism in a number of experiments involving mass-imbalanced
atomic mixtures, both for the boson-boson~\cite{Catani-2008-KRbBoseBoseSFMI}
and
boson-fermion~\cite{Ospelkaus-BoseFermi-2006,*Gunter-BoseFermi-2006,Best-RbKMixture-Feshbach-2009}
cases.

Here, we report on experimental studies of interacting one-dimensional (1D)
Bose gases in the presence of disorder. We study the effects of uncorrelated
disorder formed by atoms of an auxiliary spin state ``frozen'' to sites of an
incommensurate lattice, and compare to the case of correlated quasi-disorder
from an incommensurate bichromatic optical lattice. While both disorder types
drive strongly interacting samples into an apparent Bose glass state, a large
difference is seen for intermediate interactions, where we find that
uncorrelated disorder has a dramatic effect in driving the system towards an
insulating state. Our observation of enhanced localization for a more random
disorder is consistent with recent theoretical predictions for interacting
bosons in 1D~\cite{Roux-PRA-2008}.

\begin{figure}[b!]
    \centering
    \includegraphics[width=3.35in]{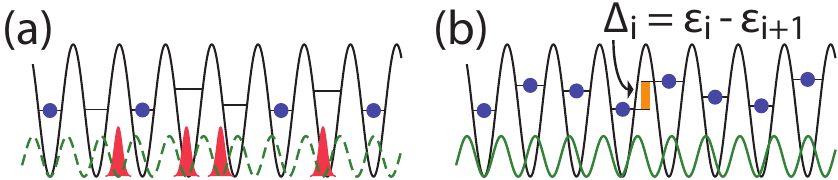}
\caption{Disordered one-dimensional Bose gases.
(a) To create a disordered impurity-field, half the atoms of a lattice-trapped Bose gas are converted to an auxiliary spin state (red) and localized to a state-selective incommensurate lattice (dashed green).
(b) Alternatively, a weak incommensurate lattice (green) is superimposed onto a lattice-trapped Bose gas.
In both cases, the secondary potential (atomic or optical) causes site-dependent energy shifts $\varepsilon_i$ and site-to-site energy differences $\Delta_i$.
}
    \label{FIG:Cartoon}
\end{figure}

The cartoon in Fig.~\ref{FIG:Cartoon}~(a) qualitatively depicts our
one-dimensional systems of lattice-trapped bosons with embedded impurities.
After creating a homogeneous spin mixture, we slowly freeze the impurity
atoms to a deep state-selective lattice of incommensurate spacing. For
comparison to a well-studied case of correlated disorder, we also study
bosons in an incommensurate bichromatic lattice
system~\cite{Fallani-TowardsBG-2007,Roati-AndersonLocalization-2008,Deissler-DisorderWithInteractions-2010},
as depicted in Fig.~\ref{FIG:Cartoon}~(b).

\begin{figure}[t!]
    \centering
    \includegraphics[width=2.75in]{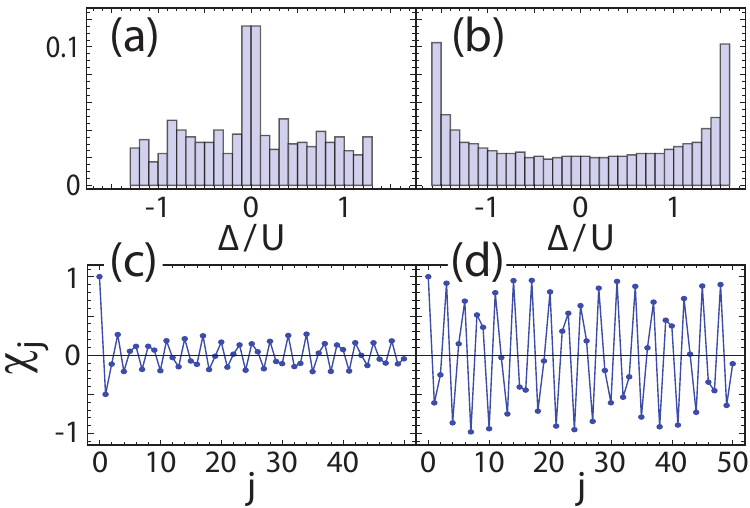}
\caption{(a,b) Histograms of calculated $\Delta$-distributions for the cases of the mixture (assuming either 0 or 1 impurities per site) and an incommensurate lattice of depth $s'=1$. Both are for a primary lattice depth $s=6$ ($U/E_R=0.4$).
(c,d) The autocorrelation function $\chi_{j}=\langle \Delta_i \Delta_{i+ j} \rangle_i / \langle \Delta_i \Delta_i \rangle_i$ of the $\Delta$-distributions in (a,b), as a function of the site-to-site distance $j$. The averaging is over $1000$ sites.
}
    \label{FIG:System}
\end{figure}

In both cases, the dynamics of the mobile atoms may approximately be
described by the Bose--Hubbard Hamiltonian
(BHH)~\cite{Fisher-Bosons-SF-IN-1989,Jaksch-ColdBosons-1998} $\hat{H}= -t
\sum_{i} (\hat{a}^{\dag}_i \hat{a}_{i+1} + \hat{a}^{\dag}_{i+1} \hat{a}_i) +
\frac{U}{2} \sum_{i}\hat{n}_{i}(\hat{n}_{i}-1) + \sum_{i} \hat{n}_{i}
\varepsilon_{i}$, where $t$ and $U$ are the tunneling and interaction
energies of the mobile atoms, and $\hat{a}_i$, $\hat{a}^{\dag}_i$, and
$\hat{n}_{i}= \hat{a}^{\dag}_i \hat{a}_i$ are the annihilation, creation, and
number operators for particles at lattice site $i$. The disordering
potentials will generally have two effects - to slightly modify the on-site
wavefunctions of the atoms and to cause random site-dependent energy shifts
$\varepsilon_i$. The first effect will lead to a spread of site-dependent
values for both $t$ and $U$ (with a multi-band treatment necessary for strong
perturbations). The second leads to random site-to-site energy differences
$\Delta_i = \varepsilon_i - \varepsilon_{i+1}$, which define the resonance
conditions for single-particle intersite tunneling. In general, the many-body
character will be defined by a competition between energy scales $t$, $U$,
and $\Delta$, with some dependence on the details of the
$\Delta$-distribution~\cite{Fisher-Bosons-SF-IN-1989}.

The two disorder potentials of Fig.~\ref{FIG:Cartoon} lead to quite different
$\Delta$-distributions. In Fig.~\ref{FIG:System} we plot calculated
histograms for (a) a fifty-percent impurity mixture and for (b) a weak
incommensurate lattice (depth $s'=1$, see experimental description below).
While both distributions are continuously filled and extend beyond $\pm
\Delta/U$, the details differ considerably. The impurity distribution is
peaked about $\Delta/U = 0$ due to adjacent impurity-free sites, while the
bichromatic lattice distribution is peaked at the outer-bounds of the
distribution. In Fig.~\ref{FIG:System}~(c,d) we plot the  normalized
autocorrelation function $\chi_j=\langle \Delta_i \Delta_{i+j} \rangle_i /
\langle \Delta_i \Delta_i \rangle_i$. The perfectly regular correlations in
Fig.~\ref{FIG:System}~(d) are a known attribute of quasi-disordered
incommensurate lattices~\cite{Roux-PRA-2008,Roscilde-Incommens-2008}. In
contrast, for the atomic impurity field - combining the irregular spacing of
bichromatic lattices and the irregular, probabilistic filling of binary
disorder - off-site correlations are strongly suppressed. This difference can
have profound effects, as the localization properties of a system will
generally depend both the strength and correlation length of the disorder
potentials~\cite{Lugan-LifshitsToBEC-2007,*Kuhn-Speckle-Transport}. An
extreme case can be found for non-interacting particles in 1D, where random
disorder leads to Anderson localization for any finite disorder strength
$\Delta \neq 0$, while incommensurate lattices induce localization only
beyond a critical lattice
depth~\cite{Giamarchi-Review-2011,*AndRefsTherein,Roati-AndersonLocalization-2008}.

To briefly describe our experimental system, we begin as
in~\cite{Gadway-BosonicLatticeMixture-2010} with an optically trapped
Bose--Einstein condensate of $^{87}$Rb atoms. In $200$~ms we load an array of
isolated, one-dimensional tubes formed by the intersection of two optical
lattices. These lattices are of period $d=\lambda/2$ and depth $40 \ E_R$
(with $\lambda = 1064$~nm, $E_R = (h/\lambda)^2 / 2 m$, Planck's constant
$h$, $m$ the atomic mass). The atoms are trapped along the tube axis $z$ by a
nearly harmonic potential of trapping frequency $\omega_z /2\pi  = 80$~Hz. A
lattice along $z$, also of period $d$ and with variable depth $s$ (in units
of the recoil energy $E_R$), is smoothly ramped up within $100$~ms. This
\emph{primary} lattice serves to define the sites (index $i$) and parameter
values ($t$, $U$) of the BHH.

\begin{figure}[b!]
    \centering
    \includegraphics[width=3.18in]{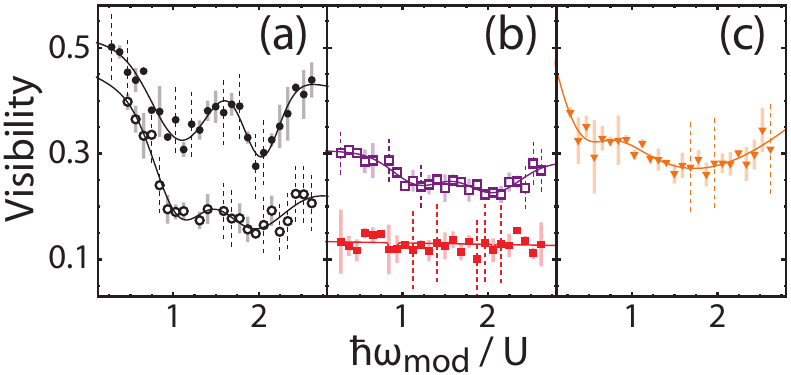}
\caption{Disappearance of excitation gap due to disorder.
(a) Visibility as a function of amplitude modulation frequency (normalized to $U / E_R = 0.53$ for $s=14$), in the absence of disorder for $s=9$ and $s=14$ (open and filled black circles). Fit lines are two Gaussians on a linear slope.
(b) For $s=14$, with atomic impurities, open purple squares and filled red squares represent $f_{\textrm{imp}}=0.1$ and $0.5$, respectively.
(c) For $s=14$, with an incommensurate lattice of depth $s' = 1$ (orange triangles) and no impurities.
Solid error bars are statistical over several runs, while dashed are estimated errors for individual runs ($120\%$ of maximum statistical error).
}
    \label{FIG:Excitation}
\end{figure}

Initially, the tubes contain only atoms in the $|F,m_F \rangle \equiv
|2,-2\rangle$ hyperfine ground state. To create atomic impurities as in
Fig.~\ref{FIG:Cartoon}~(a), a fraction ($f_{\textrm{imp}}$) of the total
population of $8 \times 10^4$ atoms is transferred to the $|1,-1\rangle$
state via a microwave Landau--Zener sweep. The impurity atoms are loaded into
a completely state-selective lattice~\cite{Gadway-BosonicLatticeMixture-2010}
along $z$ in $20$~ms. This lattice has spacing $d'=\lambda'/2$, with
$\lambda' = 785$~nm, and the impurity atoms are deeply localized at $20 \ E_R
'$ (recoil energy $E_R ' = (h/\lambda')^2 / 2 m \sim 1.8 E_R$). We typically
study an equal mixture ($f_{\textrm{imp}}=0.5$) of frozen and mobile atoms,
which we shall refer to as ``the mixture''. For the alternate disorder
implementation of Fig.~\ref{FIG:Cartoon}~(b), we begin with a sample of $4
\times 10^4$ atoms in the state $|2,-2\rangle$ (i.e. same number of mobile
atoms). We then ramp up a secondary lattice in $20$~ms, of spacing $d'$ and
variable depth $s' \times E_R'$, onto the $|2,-2\rangle$ atoms. Due to the
external trapping potential, our system is not homogeneous, but can be
characterized by a typical central filling factor of $\overline{n} \sim 3$
(total) atoms per site.

\begin{figure}[tb!]
    \centering
    \includegraphics[width=3.33in]{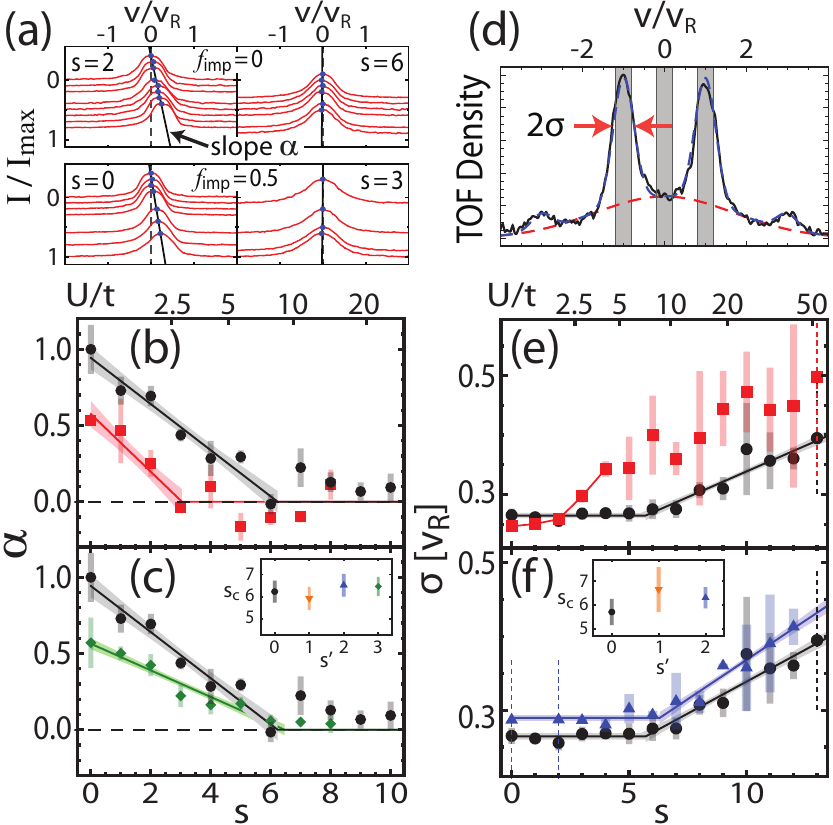}
\caption{
(a)~The response to impulse is determined by a straight-line fit (with slope $\alpha$, normalized to the case of free atoms) to the dependence of velocity on applied impulse $I$ (profiles shown for disorder-free and impurity mixture cases; recoil velocity $v_R = h/2md$).
(b)~$\alpha$ versus lattice depth $s$ for bosons without disorder (black circles) and with atomic impurities ($f_{\textrm{imp}}=0.5$, red squares). Lines and surrounding shaded regions are fits to the data with confidence regions (1~s.d.) of a linearly decaying response, with no response ($\alpha = 0$) beyond a depth $s_c$.
(c)~Similarly, but for an incommensurate lattice ($s'=3$, green diamonds), with disorder-free data reproduced for comparison. The inset shows values of $s_c$ versus $s'$ as determined by fit intercepts.
(d)~The momentum-peak width $\sigma$ ($1/\sqrt{e}$ half-width) is determined by a symmetric multi-Gaussian fit to time-of-flight interference patterns. The shaded regions about $v/v_R = 0 (\pm 1)$ are used to count $N_{0(\pm 1)}$ for visibility measurements of Fig.~\ref{FIG:Excitation}.
(e)~$\sigma$ versus $s$ for 1D bosons without disorder and for the mixture (colors/symbols as in (b)). The straight line for the disorder-free data is a fit of a linear increase beyond a depth $s_c$. The first few data points for the case of atomic impurities are connected as a guide to the eye.
(f)~Similarly, but for an incommensurate lattice with $s'=2$ (blue triangles) and with disorder-free data reproduced.
}
    \label{FIG:TransportPW}
\end{figure}

We begin our investigation into the effects of disorder by measuring
excitation spectra, which relate most directly to the distribution of
site-to-site energy shifts. Along with a finite
compressibility~\cite{Delande-compression-2009}, a gapless excitation
spectrum is a characteristic feature distinguishing a disordered Bose glass
state from a (homogenous) Mott insulator. We measure the excitation spectra
by performing amplitude-modulation
spectroscopy~\cite{Stoferle-1D-SFtoMI-2004,Fallani-TowardsBG-2007,Haller-Pinning-2010}
of the primary $z$-lattice at driving frequencies
$\omega_{\textrm{mod}}/2\pi$~\footnote{We sinusoidally modulate $s$ by $\pm
15\%$ for $80$~ms, ramp down to $s=4$ in $5$~ms, and allow $15$~ms of
thermalization. When used, the incommensurate lattice is turned on in $20$~ms
prior to modulation and off in $5$~ms concurrent with primary lattice
ramp-down. From time-of-flight interference we measure the visibility $\eta_-
/ \eta_+$~\cite{Gadway-BosonicLatticeMixture-2010}, with
$\eta_{\pm}=N_{+1}+N_{-1}\pm 2N_0$; $N_{0 (\pm 1)}$ apertures in
Fig.~\ref{FIG:TransportPW}~(d)}. For the disorder-free case ($\Delta_i
\approx 0$) in Fig.~\ref{FIG:Excitation}~(a), with the sample chosen to be
deep into the 1D Mott regime ($s=14$ , $U/t \approx 66$), the excitation
spectrum exhibits resonant structure. The resonance positions are consistent
with the excitation of particle-hole pairs at
$U/h$~\cite{Greiner-SF-MI-2002,Stoferle-1D-SFtoMI-2004} (and $2U/h$ due to
either higher-order processes or excitation at the edge of Mott
domains~\cite{Greiner-SF-MI-2002,Stoferle-1D-SFtoMI-2004}). In contrast, for
both the atomic impurity mixture (Fig.~\ref{FIG:Excitation}~(b)) and for an
incommensurate lattice of depth $s'=1$ (Fig.~\ref{FIG:Excitation}~(c)) having
comparable $\Delta$-distribution bounds, we observe flat excitation spectra
(cf.~\cite{Fallani-TowardsBG-2007}). These observations are
expected~\cite{Fisher-Bosons-SF-IN-1989} for broadly-filled
$\Delta$-distributions with bounds $\Delta_\textrm{max} > U$, and are
consistent with the system being in a Bose glass state (future
compressibility measurements~\cite{Delande-compression-2009} should allow for
the disambiguation between a true Bose
glass~\cite{Fisher-Bosons-SF-IN-1989,Krauth-SF-IN-DisorderBosons-1991} and a
disordered Mott state~\cite{Heidarian-DisorderMI-2004}).

While the observed spectral properties are consistent with Bose glass
formation, transport measurements are necessary to confirm insulating
behavior. Here, we study the effects of disorder in the transition region
between superfluid and insulator, determining the critical lattice depth at
which the systems become insulating through the study of localization and
transport. In regard to the former, the momentum-peak width of a released
sample (related to the inverse correlation length $\xi^{-1}$ of the sample
in-situ~\cite{Kollath-PeakWidth-2004}), exhibits a sudden increase
accompanying an abating superfluid fraction and loss of off-diagonal
long-range order~\cite{Trotzky-FiniteTempPhaseDiag-2010}. As for the latter,
it has been shown~\cite{Stoferle-1D-SFtoMI-2004,Fertig-1D-Damping-2005} that
the response to an applied impulse dies away upon entering the strongly
correlated regime, and can serve as a signature of insulating
behavior~\cite{Haller-Pinning-2010,Pasienski-Disorder-2010}.

To study transport, we look at how the system evolves after an applied
impulse. A magnetic-field gradient along $z$ is pulsed on for duration of
$T=1.2$~ms and applies a variable force $F$ ranging from 0 to $F_{max}/m =
1.2$~m$/$s$^2$, resulting in an impulse $I=F \times T$. As illustrated in
Fig.~\ref{FIG:TransportPW}~(a), we characterize the response as a function of
$I$ (with slope $\alpha$) by monitoring the center-of-mass velocity along $z$
in time-of-flight absorption images following a brief ($\sim$$1$~ms) ramp-off
of the $z$-lattice. We access the momentum-peak width as
in~\cite{Stoferle-1D-SFtoMI-2004,Fallani-TowardsBG-2007} by releasing the
atoms in time-of-flight following a $50$~$\mu$s lattice ramp up to $s=20$ and
a gravitational phase-shift along $z$ (without impulse or lattice ramp-off).
We then determine the peak width $\sigma$ by a fit to the profile of
symmetric diffraction peaks on top of an incoherent background, as shown in
Fig.~\ref{FIG:TransportPW}~(d).

We find that for the impurity mixture, the mobile atoms are more easily
driven towards insulating behavior. Similar to~\cite{Haller-Pinning-2010}, we
determine the critical point at which the atoms become unresponsive to
impulse by fitting a linear decay to the response $\alpha$ as a function of
lattice depth. As shown in Fig.~\ref{FIG:TransportPW}~(b), this fit to the
transport measurements yields a critical depth $s_c = 3.0 \pm 0.5$. While we
do not fit the peak width data, a kink near $s_c \approx 2$ can be observed
in Fig.~\ref{FIG:TransportPW}~(e). These values are roughly half of those
measured for the disorder-free case, with values of $s_c = 6.2 \pm 0.5$ and
$5.7 \pm 0.5$ based on impulse response and peak width, respectively. This
value of $s_c \approx 6$ is close to our expectations based on the mean
Lieb--Liniger parameter~\footnote{$\gamma = m g_{\textrm{1D}} / \hbar^2
n_{\textrm{1D}}$~\cite{Giamarchi-Review-2011,*AndRefsTherein}, with 1D
density $n_{\textrm{1D}}$ and effective 1D interaction strength
$g_{\textrm{1D}} \approx 2 \hbar \omega_\perp
a$~\cite{Olshanii-1Dscattering-1998}, s-wave scattering length $a \approx
5.3$~nm, and transverse trapping frequency $\omega_\perp = 2\pi \times
26$~kHz} ($\gamma = 0.6$) of our 1D gases prior to lattice-loading, with an
estimate based on the Bose--Hubbard (Sine--Gordon) model predicting $s_c =
6.2$~($4.7$)~\cite{Haller-Pinning-2010}.

For the incommensurate lattice, transport data for $s'=3$ is shown in
Fig.~\ref{FIG:TransportPW}~(c). In this case no shift of the critical depth
is seen, and as shown in the inset, this is the case for all incommensurate
lattice depths considered ($s' \leq 3$, where we restrict to $s'/s < 1$ in
the transition region to maintain the perturbative nature of the
incommensurate lattice). The momentum-peak width data mirrors this lack of a
shift of the transition point (inset: for all depths considered $s' \leq 2$).

While both incarnations of disorder resulted in an apparent Bose glass state
for very deep lattices, a clear difference was seen in their effect on more
weakly interacting samples. In attempting to account for the observed
difference, a natural consideration is the disparity in their correlations
[$\chi_j$,~cf.~ Fig.~\ref{FIG:System}~(c,d)]. In general, one expects that
the less correlated the disorder, the more enhanced is the
localization~\cite{Lugan-LifshitsToBEC-2007,*Kuhn-Speckle-Transport}. For
interacting bosons in 1D, it has been shown
theoretically~\cite{Roux-PRA-2008} that the localization transition occurs
for a more weakly interacting gas (larger values of the Luttinger exponent
$K$ or lower values of the Lieb--Liniger parameter
$\gamma$~\cite{Giamarchi-Review-2011,*AndRefsTherein}) in uncorrelated
disorder than for correlated disorder. Our observations of a sizeable shift
of the transition point for impurities and a negligible shift for an
incommensurate lattice are thus in qualitative agreement with expectations
based on their dissimilar correlation properties.

Also relevant to our observations is the reduction of phase-space density in
the presence of localized impurities, as well as of the atomic density due to
the dynamical formation of impurities from the mobile species. The first
effect has been
shown~\cite{Snoek-Bloch-2011,*Cramer-InteractionDependentTempEffects-2011} to
be responsible for adiabatic heating and loss of coherence in recent
Bose--Fermi mixture experiments~\cite{Best-RbKMixture-Feshbach-2009}, due to
reduced entropy following a reduction in effectively occupiable sites, both
for attractive and repulsive interactions. The second effect is more
particular to ``quantum
emulsion''~\cite{RoscildeQuantumEmulsion-2007,Horstmann-PRL-2010,Gadway-BosonicLatticeMixture-2010}
experiments. Here, reduced density leads to more strongly correlated
many-body states in 1D and thus favors increased localization and insulating
behavior.

In conclusion, we have observed signatures of Bose glass formation in 1D Bose
gases with superimposed disorder, both for atomic impurities and for
quasi-disordered bichromatic lattices. The two disorder types have
dramatically different effects in the transition region between superfluid
and insulator, with atomic impurity disorder inducing localization in much
more weakly interacting gases. Our observation that a more weakly correlated
disorder leads to enhanced localization is in qualitative agreement with
recent theoretical predictions for interacting 1D boson systems.

We thank Martin G. Cohen for valuable comments on the manuscript. This work
was supported by NSF (PHY-0855643) and the Research Foundation of SUNY. B.G.
and J.R. acknowledge support from the GAANN program of the US DoEd.



\end{document}